\documentstyle[aps,pre,epsf,preprint]{revtex}
\include{epsf}

\begin{document}

\title{Diffusion at constant speed in a model phase space} 

\author{S. Anantha Ramakrishna\footnote{Presently at the Blackett Laboratory, 
Imperial College, London SW7 2BZ, U.K.} and N. Kumar}
\address{Raman Research Institute, C.V. Raman Avenue, Bangalore 560 080, India}
\date{\today}
\maketitle

\begin{abstract}
	We reconsider the problem of diffusion of particles  at constant speed
and present a generalization of the Telegrapher
process to higher dimensional stochastic  media ($d>1$), where the particle can 
move along $2^d$ directions. We derive the
equations for the probability density function using the 
``formulae of differentiation'' of Shapiro and Loginov.
The model is an advancement over similiar models of photon migration 
in multiply scattering media in that it results 
in a true diffusion at constant speed in the limit of large dimensions. 
\end{abstract}

\section{Introduction}
Diffusion theory is an old subject and has been applied with enormous success
in many fields of Physics, Chemistry and Biology. But pure diffusion of 
particles is actually unphysical as it neglects the inertia of the particles. 
It is a Wiener process for the spatial 
position of the particles and holds strictly in the limit when the trasnport 
mean free path $l^{*} \rightarrow 0$ and the speed $c \rightarrow \infty$ such
that the diffusion coefficient $D=cl^{*}/3$ is a constant. In other words the 
particle scatters at every point and has no directional memory or persistence. 
The persistence becomes important in many problems when processes at short 
length and time scales become important as in polymer physics and radiative 
transport in random media. The incoherent energy transport of light in random
media described by the radiative transfer equation~\cite{chandra50,ishimaru} can
be approximated as a diffusion process in a coarse grained description. 
But again, the diffusion approximation fails here at short length- and 
time-scales\cite{alfano90}, 
not because of any finite photon mass, but due to the anisotropic 
(forward) scattering nature of the finite size scatterers. 
This becomes particularly 
important in the biomedical imaging and diagnostic applications with NIR light
\cite{tuchin} when the ballistic and `snake' (near ballistic) light 
in highly forward scattering media are
used. Since the general analytic solutions to the radiative transfer equation 
that take into account the full scattering properties are 
not known even for simple geometries, it becomes important to develop simple
models that incorporate the physical persistence of the problem into the 
stochastic process. Further in the case of light, assuming the distance between the scatterers is much larger than the wavelength of the light, 
the speed of the photon is 
constant in between the scattering events. While approximately describing 
light as a particle undergoing stochastic motion, this constancy of the speed 
should be preserved at the very least.  

Recently we successfully described this diffusion at a constant 
speed of photons as a diffusion on the velocity sphere~\cite{sar_pre}.
Among the other
models proposed to deal with the intermediate range of length- and time-scales
between the ballistic motion and diffusive transport, the attempts to generalize
the Telegrapher equation are important. The Telegrapher equation is exact in one
dimension and describes the diffusion of a particle whose speed is fixed, i.e.,
the velocity can take on only two values $\pm c$ \cite{goldstein}.  The
Telegrapher equation, i.e,
\begin{equation}
\frac{\partial^{2} P}{\partial
t^{2}} + \Gamma \frac{\partial P}{\partial t} -  c^{2}
\frac{\partial^{2} P}{\partial x^{2}} = 0   ,
\end{equation}
where P is the probability distribution function and
$\Gamma /2$ is the mean scattering rate, is a combination of the wave 
equation describing the inertial (persistent) aspect 
and of the diffusion equation 
describing the stochastic aspect. This equation has found wide application 
in many fields~\cite{maddox} and was first considered by J.C. Maxwell%
\cite{maxwell,repphys}, 
more than a century ago in his attempt to describe heat conduction from 
basic kinetic theory.  He discarded the inertial term, however, on the grounds
that it would be important only at extremely short time scales. 
It also has since been shown to describe the 'second 
sound' in liquid Helium II, and classical mesoscopic diffusion obeying the 
Maxwell-Cattaneo law instead of the usual Fick's law\cite{cattaneo}.  

Following a suggestion of Ishimaru~\cite{ishimaru_ao}, a
generalization of the Telegrapher equation to higher dimensions in a heuristic 
manner was attempted~\cite{durian_josa97} by simply
replacing $\partial^{2} P / \partial x^{2} $  in Eqn.(7.1) by $\nabla^{2} P$,
to describe
photon migration at short length scales, by including some ballistic aspects.
This {\it ad-hoc} generalization appeared to be quite successful to 
describing photon migration as it preserved causality and did much better 
than the diffusion approximation to describe photon transport in absorbing 
media \cite{durian_josa97,durian_ol98}. This was also extended to
studies of Diffusing Wave Spectroscopy in thin samples \cite{durian_pre98}.  
It was,
however, shown by comparing it with Monte-Carlo simulations that this
generalization furnished no better an approximation than the diffusion
approximation in higher dimensions \cite{porra_pre97}.
The in-principle weakness of this `ad-hoc' approach was demonstrated 
by the fact that the photon probability density evolving under 
this equation becomes negative
 in two dimensions for the simplest case of an unbounded (infinite) 
medium at short times ($t \sim t^{*}$) when the ballistic aspects 
of transport are most important. In fact, the negativity of the solution 
to this equation is a generic property
in even-dimensional spaces \cite{porra_pre97,morse}. 

In this paper, we reconsider the problem of diffusion of photons at constant
speed and present a generalization of the
Telegrapher process to higher dimensional turbid media ($d>1$),
where the photon can move along $2^{d}$ directions along the diagonals of a 
$d$-dimensional hypercube. We derive the equation
for the probability density function using the ``formulae of differentiation''
of Shapiro and Loginov \cite{shapiro}, by considering a correlated random walk at constant speed. We show that a partial differential equation of
order $2^{d}$ results for the probability distribution function in
$d$-dimensions. Our model is an advancement over the earlier models of Bogu\~{n}\'{a
} {\it et al.} \cite{boguna_pre98,boguna_pre99}, where the  photon could only 
move  along the $2d$ directions  along the axes, and  results in a true 
diffusion at constant speed in the limit of large dimensions. 
Our work brings out certain features that were not recognized in earlier work. 
Light in the stochastic medium is considered to be a particle on
which the medium exerts fluctuating forces.  Each scattering event only changes
the direction of the photon without affecting the speed of propagation.

\section{ The Telegrapher process in one dimension}
Let us first consider the dynamics of a particle executing a random walk in one
dimension while moving with constant speed $c$.  This would describe the motion
of light in a disordered fibre, or of electrons on the Fermi points in a
 one-dimensional
disordered wire, if we neglect the wave nature and the consequent Anderson
localization (Strictly speaking, this description would not hold for 1-D where
all the quantum states are localized states.  For sample lengths much smaller than
the localization length, however, the transport is almost diffusive).  The
velocity $v(t)$ of the particle is a random function of time such that it can
take only two values $\pm c$, i.e.,  a Dichotomic Markov process.  If $\Gamma
/2$ be
the transition probability per unit time between these two values of the
velocity ($\langle ~~\rangle$ indicate averaging over the disorder),
\begin{eqnarray}
\langle v(t) \rangle & = & 0 ,\\
\langle v(t) v(t') \rangle & = &
c^{2} \exp (-\Gamma \vert t - t'\vert ),
\end{eqnarray}
i.e.  the velocity is
exponentially correlated in time.  We note that the stochastic Langevin
equation for the displacement $\dot{x} = v(t)$ gives:
\begin{eqnarray} \langle x \rangle & = & 0, \\
\langle x^{2} \rangle & = & \frac{2c^{2}}{\Gamma} ( t 
- \frac{1-e^{-\Gamma t}}{\Gamma}),
\end{eqnarray}
i.e.,  the behaviour at long times
($t \rightarrow \infty$) or very large scattering strengths (large $\Gamma$) is
diffusive ($\langle x^{2} \rangle \sim t$) and at short times ($ t \rightarrow 0
$), the behaviour is ballistic ($\langle x^{2} \rangle \sim t^{2}$).  

Next we will derive the equation for the probability distribution function. Let
$\Pi(x;t)$ be the phase space density of points in the x-t phase space.  Now,
$\Pi$ satisfies the Stochastic Liouville equation :
\begin{equation}
\frac{\partial \Pi}{\partial t} + \frac{\partial }{\partial x}(\dot{x} \Pi) = 0.
\end{equation}
 Averaging over all the realizations of the random function
$v(t)$, by the van Kampen lemma \cite{vankampen}, the probability distribution
$P(x;t) = \langle \Pi(x;t) \rangle$.  We also define $W(x;t) = \langle v(t)
\Pi(x;t) \rangle$ and obtain,
\begin{equation} \frac{\partial P}{\partial t} +
\frac{\partial W}{\partial x} = 0 .
\end{equation} Now using the ``formulae of
differentiation'' of Shapiro and Loginov \cite{shapiro}, for a Dichotomic Markov
process,
\begin{equation} \frac{\partial W}{\partial t} = -\Gamma W - c^{2}
\frac{\partial P}{\partial x}.
\end{equation}
Eliminating W from the above
equations, we obtain
\begin{equation}
\label{tge1d}
\frac{\partial^{2}P}{\partial t^{2}}
+ \Gamma \frac{\partial P}{\partial t} = c^{2}
\frac{\partial^{2} P}{\partial x^{2}} ,
\end{equation} i.e.  the Telegrapher
equation for the probability distribution function $P$.  This is an exact
description of the motion of a particle with constant speed in 1-D.

The solutions to Equation.(\ref{tge1d}) are well known and, in an infinite 
medium given by \cite{goldstein,durian_josa97} 
\begin{equation}
P(x,t;x=0,t=0) = \left( \frac{\Gamma}{2c} \right)e^{-\Gamma t/2} \big[ I_{0}(y) 
+2\Gamma t \frac{I_{1}(y)}{y} \big] ~\theta(ct- \vert x\vert ),
\end{equation}
where $ y = (\Gamma/2c) \sqrt{c^{2}t^{2} - x^{2}}$, $I_{0}$ and $I_{1}$ are
the modified Bessel functions of order zero and one respectively, and $\theta$
is the Heaviside step function. Note that the solution is zero for $\vert x\vert > c t$ and thus causality is preserved. The solution indicates that the particles spread out symmetrically 
from the origin, half of them to the left and the other half to the right, 
with a front velocity of $c$ beyond which there are no more particles and 
heap up at the `light fronts' where there are $N/2 \exp(-\Gamma t/2)$ particles 
(if there are $N$ particles altogether).
 
\section{ Generalization to higher dimensions}
\begin{figure}
\epsfxsize=400pt
\begin{center}{\mbox{\epsffile{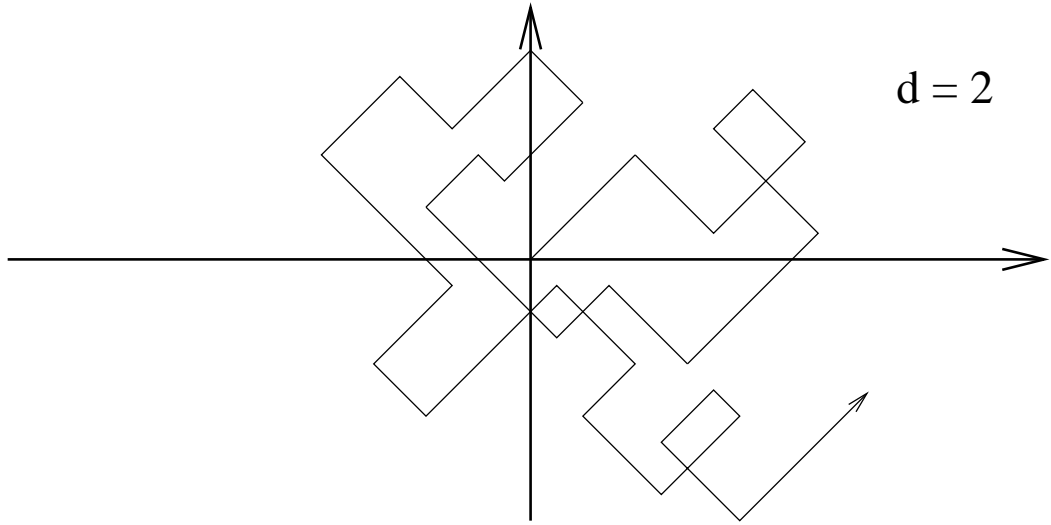}}}
\end{center}
\caption{ One realization of the possible real space trajectories for the 
generalized Telegrapher process in two dimensions.
\label{modelspace}}
\end{figure}

Now, we seek a generalization of the Telegrapher process process to higher 
dimensions. The simplest way of doing this  is to make every
orthogonal component a
dichotomic Markov process which take the values $\pm c$. Thus, in 
$d$ dimensions, the particle is seen to move along the
diagonals of the $d$ dimensional hypercube with a speed $\sqrt{d} c$.
Here the
real space is continuous, while the velocity space is discrete and can take
only $2^{d}$ discrete values in $d$ dimensions. This is the model phase space 
that we consider the photon to execute a random walk in. 
For example, in Fig. \ref{modelspace}, 
we show a possible trajectory in the real space for $d=2$ and a particle 
released from the origin at $t=0$.
\pagebreak

\subsection{The equations for the generalized Telegrapher process}
In two dimensions, we will consider both the x and y components of the velocity of the
particle to be independent dichotomic Markov processes,
\begin{eqnarray}
v_{x}(t) & = & c \chi_{1}(t), \\ v_{y}(T) & = & c \chi_{2}(t), \\
\langle \chi_{i}(t) \rangle & = & 0, \\ \langle \chi_{i}(t) \chi_{j}(t') \rangle
 & = &
\delta_{ij} \exp (-\Gamma \vert t -t' \vert),
\end{eqnarray}
where $\chi_{i}(t)$ are unimodular processes.  The particle is thus seen to move
along the four directions $(\pm \hat{i},\pm \hat{j})$ with a constant speed
$\sqrt{2}c$. Now using the stochastic Louiville
equation, We define the averages $P(x,y;t) = \langle \Pi(x,y;t) \rangle$,
$W_{x} = \langle v_{x}(t) \Pi(x,y;t) \rangle$, $W_{y} = \langle v_{y}(t)
\Pi(x,y;t) \rangle$ and $W_{xy} = \langle v_{x}(t) v_{y}(t) \Pi(x,y;t)
\rangle$.  We generalize the ``formula of differentiation'' of Shapiro and
Loginov to n-independent dichotomic Markov processes
as \begin{eqnarray}
\frac{d}{dt} & \langle & v_{1} v_{2} \cdots v_{n} \Pi [ v_{1}, v_{2}, \cdots
v_{n}] \rangle_{v_{1}, v_{2}, \cdots , v_{n}} = \nonumber \\ & = & -(\Gamma_{1}
+ \Gamma_{2} + \cdots + \Gamma_{n}) \langle v_{1} v_{2} \cdots v_{n} \Pi [
v_{1}, v_{2}, \cdots , v_{n}] \rangle_{v_{1}, v_{2}, \cdots v_{n}} 
\nonumber \\ & + &\langle
v_{1} v_{2} \cdots v_{n}\frac{d\Pi [ v_{1}, v_{2}, \cdots
v_{n}]}{dt}\rangle_{v_{1}, v_{2}, \cdots , v_{n}},
\end{eqnarray}
where $v_{i}(t)$ are independent dichotomic Markov processes, $\Gamma_{i}$ are 
the
respective transition rates and $\Pi$ is a functional of $v_{1}$, $v_{2}$,
$\cdots $, $v_{n}$.  Using the above, we obtain the following closed set of
equations :
\begin{eqnarray} 
\label{tgrapheq2d}
\frac{\partial P}{\partial t} + \frac{\partial
W_{x}}{\partial x} + \frac{\partial W_{y}}{\partial y} & = & 0 ,\\
\frac{\partial
W_{x}}{\partial t} + \Gamma W_{x} & = & -c^{2} \frac{\partial P}{\partial x} -
\frac{\partial W_{xy}}{\partial y} ,\\
\frac{\partial W_{y}}{\partial t} + \Gamma
W_{y} & = & -c^{2} \frac{\partial P}{\partial y} - \frac{\partial
W_{xy}}{\partial x} ,\\
\frac{\partial W_{xy}}{\partial t} + 2 \Gamma W_{xy} & =
& -c^{2} \left[ \frac{\partial W_{y}}{\partial x} + \frac{\partial
W_{x}}{\partial y} \right].
 \end{eqnarray}
 Eliminating $W_{x}, W_{y}, W_{xy}$
from the above set of equations, we obtain for the probability distribution
function $P(x,y;t)$ :
\begin{equation} \frac{\partial}{\partial t} \left(
\frac{\partial}{\partial t} + 2 \Gamma \right) \left( \frac{\partial}{\partial
t} + \Gamma \right)^{2} P - 2 c^{2} \left( \frac{\partial}{\partial t} + \Gamma
\right)^{2} \nabla^{2} P + c^{4} \left( \frac{\partial^{2}}{\partial x^{2}} -
\frac{\partial^{2}}{\partial y^{2}} \right) P = 0 ~~.
\label{twodtelegrapheq}
\end{equation}
By performing a $\pi /4$ rotation of the space axes and rescaling the speed to
$c$, this equation is seen to be the same as the one derived by Bo\~{g}una et
al.\cite{boguna_pre98} by a different approach.  Unlike the ``{\it ad-hoc}
Generalized Telegrapher equation'' of Durian and Rudnick \cite{durian_josa97}, 
this partial differential equation is of fourth order involving all space and 
time derivatives.  

Similiarly, in three dimensions, we consider the x, y and z components to the
velocity to be independent dichotomic Markov processes.  Again, following the
above procedure, we obtain the closed set of eight equations :
\begin{eqnarray}
\label{tgrapheq3dfirst}
\frac{\partial P}{\partial t} + \frac{\partial W_{x}}{\partial x} + \frac{
\partial W_{y}}{\partial y}  + \frac{\partial W_{z}}{\partial z} & = & 0 \\
\frac{\partial W_{x}}{\partial t} + \Gamma W_{x} & = & -c^{2} \frac{\partial P}{
\partial x} -  \frac{\partial W_{xy}}{\partial y} - \frac{\partial W_{zx}}{
\partial z}\\
\frac{\partial W_{y}}{\partial t} + \Gamma W_{y} & = &-c^{2} \frac{\partial P}{
\partial y} -  \frac{\partial W_{xy}}{\partial x} - \frac{\partial W_{yz}}{
\partial z}\\
\frac{\partial W_{z}}{\partial t} + \Gamma W_{y} & = &-c^{2} \frac{\partial P}{
\partial z} -  \frac{\partial W_{zx}}{\partial x} - \frac{\partial W_{yz}}{
\partial y}\\
\frac{\partial W_{xy}}{\partial t} + 2 \Gamma W_{xy} & = & -c^{2} \left[ \frac{
\partial W_{y}}{\partial x} +  \frac{\partial W_{x}}{\partial y} \right] - \frac{
\partial W_{xyz}}{\partial z} \\
\frac{\partial W_{yz}}{\partial t} + 2 \Gamma W_{yz} & = & -c^{2} \left[ \frac{
\partial W_{y}}{\partial z} +  \frac{\partial W_{z}}{\partial y} \right] - \frac{
\partial W_{xyz}}{\partial x} \\
\frac{\partial W_{zx}}{\partial t} + 2 \Gamma W_{zx} & = & -c^{2} \left[ 
\frac{\partial W_{z}}{\partial x} +  \frac{\partial W_{x}}{\partial z} \right] - \frac{
\partial W_{xyz}}{\partial y} \\
\label{tgrapheq3dlast}
\frac{\partial W_{xyz}}{\partial t} + 3 \Gamma W_{xyz} & = & -c^{2} \left[ \frac
{\partial W_{yz}}{\partial x} +  \frac{\partial W_{zx}}{\partial y} + \frac{
\partial W_{xy}}{\partial z} \right]
\end{eqnarray}
It is possible to obtain a cumbersome-looking partial differential equation for
P(x,y,z;t) alone, similiar to Equation(\ref{twodtelegrapheq}) 
by eliminating the other functions
from the above coupled set of differential equations.  However, no extra
information results and it will not be presented here.  

 The partial differential equation for $P(\vec{r};t)$ alone, in general, will be
of order $2^{d}$ (there are $2^{d}$ independent first order coupled differential
equations for $d$ dimensions).
The set of coupled first order differential
equations(\ref{tgrapheq3dfirst}-\ref{tgrapheq3dlast}) offer a very convenient factorization of the $2^{d}$
dimensional equation satisfied by $P(\vec{r};t)$ and are a more convenient
starting point for numerical calculations of the solutions.
We note that the
above equations are linear with constant coefficients and can easily be solved
by taking Laplace transforms with respect to time and space variables.
Inverting the solutions obtained back to the real space-time, however, is
non-trivial and only a few characteristic quantities such as the moments of the
residence times have been calculated \cite{boguna_pre99} for a similiar model. 

\subsection{Absorbing boundary conditions}
For the above system of equations, absorbing boundary conditions can be easily and rigorously applied for this set
of equations. For an absorbing boundary at $x=0$, with the stochastic medium
occupying the negative semi-infinite half-space, the appropriate boundary conditions
corresponding to no incoming flux are  
\begin{eqnarray}
W_{x}(x=0,y,z;t) = c P(x=0,y,z;t), \\
W_{xy}(x=0,y,z;t)= c W_{y}(x=0,y,z;t),\\
W_{zx}(x=0,y,z;t)=W_{z}(x=0,y,z;t), \\
W_{xyz}(x=0,y,z;t)= cW_{yz}(x=0,y,z;t),
\end{eqnarray}
and free boundary conditions on the
other functions $P$, $W_{y}$, $W_{z}$, $W_{yz}$. These conditions are equivalent
to the integral boundary condition $\partial/ \partial t \int_{-\infty}^{x=0}
P(x,y,z;t) dx = c P(x=0,y,z;t)$ on $P(x,y,z;t)$ alone. \\

\subsection{Projected motion along any axis and angular non-symmetry of the model}
In higher dimensions ($d > 1$), the {\it ad-hoc}
generalized Telegrapher equation {\it viz.}
$\frac{\partial^{2} P}{\partial t^{2}}
    +  \Gamma \frac{\partial P}{\partial t} - c^{2} \nabla^{2} P = 0 $
is indeed obtained only if higher
order velocity correlations are neglected, {\it i.e.},  terms such as $W_{xy} =
\langle v_{x} \Pi \rangle$, $W_{xyz}$ etc. are set to zero.  This is not
correct especially at short times, when we expect the velocity components to be
correlated to a quite some extent. However, it is easily seen that the
marginal probability distribution for  the projected motion along one of
the axis $ p(x_1;t) = \int P(x_{1},x_{2},\cdots,x_{d}) dx_{1}~dx_{2}~\cdots~dx_{
d}$
satisfies the Telegrapher equation  $\frac{\partial^{2} p}{\partial
t^{2}} +  \Gamma \frac{\partial p}{\partial t} - c^{2} \frac{\partial^{2} p}{
\partial
x^{2}} = 0 $.
The partial differential equation for $P(\vec{r};t)$ alone, in general, will be
of order $2^{d}$ (there are $2^{d}$ independent first order coupled differential
equations for $d$ dimensions) corresponding to $2^{d}$ directions.

There are some subtle
differences between our model and that of Bogu\~{n}\'{a} et al\cite{boguna_pre98}. 
First of all, the number of allowed directions for the photon motion is greater
 in our model ($2^{d}$) than theirs ($2d$).
The reason is that, they consider that the motion of the particle
to be  along the axes, while in our case, the motion
is along the diagonals of the $d$ dimensional hypercube.  They do not obtain 
a Telegrapher equation for the marginal probability distribution for the 
projected motion in general as we do. We always have a
Telegrapher process along any one axis.  This can be best compared in two
dimensions by carrying out a $\pi/4$ rotation of the axes in Eqns.(16-19) and
then looking at the projected motion along the axes.  We obtain the Telegrapher
equation $\frac{\partial^{2} p}{\partial t^{2}} + 2\Gamma \frac{\partial
p}{\partial t}-c^{2}\frac{\partial^{2} p}{\partial x^{2}} = 0 $ , {\it i.e.},
only the diffusion coefficient is renormalized.  This corresponds to the three
step Telegrapher process of moving at constant speed to the left or the right
with a probability $1/4$ and being at rest with a probability $1/2$.
In higher dimensions ($d >2$),
the diagonals of the hypercube are not orthogonal and the equation obtained for
the projected motion along the diagonals is not a Telegrapher
equation in our case.

Thus, it is to noted that in these models
without angular symmetry, such a description of projected motion is
non-unique and depends on the direction of the projected motion. It should,
however, be pointed out that the angular spacing between these discrete
directions, given by the ratio of the total solid angle to the number
of directions, in our model is
\begin{equation}
\frac{\Omega(d)}{2^{d}} = \frac{1}{2^{d}} \frac{2 \pi^{d/2}}{\Gamma(d/2)~d}
\rightarrow \left( \frac{\pi^{1/2} e^{1/2}}{2 \ln^{1/2}(d/2)} \right)^{d},
\nonumber
\end{equation}
where $\Gamma$ here is the Gamma function. In the limit of large $d$, the
angular spacing  decays almost exponentially to zero. Thus, in the limit of
large dimensions, this process indeed describes a genuine
diffusion at constant speed. This result is due to the exponential dependence
on the dimensionality for the number of allowed directions for the 
photon velocity and is not obtained
by  Bogu\~{n}\'{a} et al.  

\section{Kubo-Anderson like stochastic processes}
In passing, we would like to touch upon the possible generalization to more
complex stochastic processes, which  demonstrates the power of the
current approach. It is a simple matter now, to write down the equations for
probability distribution function for a Kubo-Anderson like
process, given by the sum of n-independent dichotomic Markov processes
(in one -dimension), i.e.,
$v(t) = v_{1}(t)+v_{2}(t)+ \cdots +v_{n}(t)$, where $v_{i}(t)$ can take on the
values $\pm c$ and $\langle v_{i}(t) \rangle = 0;~~ \langle v_{i}(t)v_{j}(t')
\rangle =
c^{2} \exp(-\Gamma \vert t-t' \vert ) \delta_{ij}$. The structure of the
equations for $n=2$ and $n=3$ remain the same as Eqn. (16-19)
 and Eqn. (21-28) respectively, with only the derivatives $\partial /\partial
y$ and $ \partial /\partial z$ both replaced by $\partial /\partial x$. It is
immediately seen that the case of $n=2$ corresponds to the 3-step Telegrapher
processes described above. The generalization to higher $n$ is obvious. In
general, one obtains $n+1$ coupled partial differential equations
for the sum of $n$-independent dichotomic Markov processes. 

\section{Conclusions}
In conclusion, we have developed a particular `generalization' of the
Telegrapher process
to higher dimensional ($d>1$) stochastic media which could be potentially
useful for studying photon migration in turbid media as it rigourously
preserves the photon speed to be constant between the scattering events.
In comparision to the model presented in Ref.\cite{sar_pre}, 
where the photon's random
walk was modelled as a diffusion on the velocity sphere, we have a model phase 
space here, where the photon can move only along the $2^{d}$ directions of the 
diagonals of the $d$ dimensional hypercube. It is admittedly an  artificial 
phase space, but one having an appreciable directional persistence unlike 
the zero-persistence diffusion theory. However, the model does not 
have angular symmetry. In 1-D, there are only two directions and 
hence, the Telegrapher equation is  
exact. On the other hand, in the 
limit of very large dimensions, the angular spacing between the 
directions tends to zero almost
exponentially, and again the process is indeed  a genuine diffusion-at-a-constant-speed.
It has been shown that the equation for the projected motion along any 
hypercube axis is a 1-D Telegrapher equation, though it is non-invariant under
an arbitrary  rotation. Further, the {\it ad-hoc} generalized 
Telegrapher equation in higher
dimensions \cite{durian_josa97,durian_ol98} is recovered when higher order
correlations are neglected. The power of this approach is demonstrated by
deriving the equations for a sum of n-independent Markov processes.

\end{document}